\documentclass[aps,prb,floatfix,twocolumn,showpacs,superscriptaddress,groupedaddress]{revtex4-1}  
\usepackage{graphicx}  
\usepackage{amsmath,amssymb}
\usepackage{flushend}
\usepackage[colorlinks = true,
            linkcolor = blue,
            urlcolor  = blue,
            citecolor = blue,
            anchorcolor = blue]{hyperref}
\usepackage{array}
\usepackage{float}
\usepackage{natbib}
\usepackage{booktabs}
\mathchardef\Re="023C
\mathchardef\Im="023D

\usepackage{color}
\usepackage{times}

\begin{document}

\title{Interplay of Disorder and Point-Gap Topology: Chiral Modes, Localization and Non-Hermitian Anderson Skin Effect in One Dimension}

\author{Ronika Sarkar} \email[]{ronikasarkar@iisc.ac.in}\affiliation{Department of Physics,
Indian Institute of Science, Bangalore 560012, India.}
\author{Suraj S. Hegde} \email[]{suraj.hegde@tu-dresden.de}
\affiliation{Institute of Theoretical Physics and Wurzburg-Dresden Cluster of Excellence ct.qmat, Technische Universitat Dresden, 01069 Dresden, Germany.}
\author{Awadhesh Narayan} \email[]{awadhesh@iisc.ac.in} \affiliation{Solid State and Structural Chemistry Unit,
Indian Institute of Science, Bangalore 560012, India.}

\vskip 0.25cm

\begin{abstract}

Symmetry-protected spectral topology in extended non-Hermitian quantum systems has interesting manifestations such as dynamically anomalous chiral currents and skin effect. In this work, we study the interplay between symmetries and disorder in a paradigmatic model for spectral topology - the non-reciprocal Su-Schrieffer-Heeger model. We consider the effect of on-site perturbations (both real and purely imaginary) that explicitly break the sub-lattice symmetry. Such symmetry-breaking terms can retain a non-trivial spectral topology but lead to a different symmetry class. We numerically study the effect of disorder in on-site and non-reciprocal hopping terms. Using a real-space winding number that is self-averaging and quantized, we investigate the impact of disorder on the spectral topology and associated anomalous chiral modes under periodic boundary conditions. We discover a remarkable robustness of chiral current and its self-averaging nature under disorder. The value of the chiral current retains the clean system value, is independent of disorder strength and is  tracked completely by the real-space winding number for class A which has no symmetries, and class AIII, which has a sub-lattice symmetry. In class $D^\dagger$, which has $PT$-symmetric on-site gain and loss terms, we find that the disorder-averaged current is not robust while the winding number is robust. We study the localization physics using the inverse participation ratio and local density of states. As the disorder strength is increased, a mobility-edge phase with a finite winding appears. The abrupt vanishing of the winding number marks a transition from a partially localized to a fully localized phase. Under open boundary conditions, we similarly observe a series of transitions through skin effect--partial skin effect--no skin effect phases. Further, we study the non-Hermitian Anderson skin effect (NHASE) for different symmetry classes, where the system without skin effect develops a disorder-driven skin effect at intermediate disorder values. Remarkably while  NHASE is present for different classes, the real-space winding number shows a direct correspondence with it only when all symmetries are broken.

\end{abstract}

\maketitle

\date{\today}

\section{Introduction}

Non-Hermitian systems, both in classical and quantum settings, have gained interest because of their physical relevance and conceptual novelty, \cite{2,59,60,61,40, ronny1, ronny2,Li_2022}. Non-Hermitian concepts find a wide range of applicability in fields such as optics, photonics, acoustics, mechanical metamaterials, and biological open systems, to name a few -- rapid experimental work is being done to explore non-Hermitian aspects in these platforms \cite{40,41,42,71,Ghatak_2020}. Non-Hermitian lattice models have gained much importance in condensed-matter settings, having introduced many interesting ideas such as complex eigen-spectra, non-Hermitian skin effect \cite{8,30,73,51,76,78,79,80,81,82,83,84,https://doi.org/10.48550/arxiv.2205.10379,Borgnia_2020,PhysRevA.104.022215,Li_2021}, complex energy gaps, broken bulk-boundary correspondence \cite{13,14,85,86,87,88}, exceptional points \cite{32,92,93,94,95,96,97,98,99,PhysRevLett.118.045701} among others. In contrast to the band topology in Hermitian systems, the complex energy spectrum of non-Hermitian Hamiltonians is known to have a notion of topological winding, which we term `spectral topology'. Skin effect and chiral modes, among others, are manifestations of this spectral topology. Much like band topology, it is protected by anti-unitary symmetries and has been thoroughly classified under Bernhard-Sinclair classification scheme \cite{1,113,18,114,115}. The spectral topology and the accompanying features such as the chiral modes and currents have already been probed in experiments using photonic lattices \cite{Wang_2021,56,58}.\\

On the other hand, disorder has always played a vital role in condensed matter systems, ranging from the seminal works on Anderson localization \cite{36,52,53,54,55} to several works on disorder-driven phase transitions in topological systems \cite{37,38,39, 63,64,69,74,Jiang_2022}. The notion of non-Hermiticity and its interplay with disorder has been a recent subject of wide ranging interest. The effect of disorder on the inherently non-Hermitian Hatano Nelson model \cite{28,108,109,110,111,112} has been studied recently, where disorder has been introduced in the hopping to obtain interesting physics, such as disorder driven phase transitions from an extended phase to a bulk localized phase \cite{7} and an  non-Hermitian Anderson skin effect, where disorder gives rise to a skin effect \cite{9}, and more. Some studies of the non-Hermitian Su-Schrieffer-Heeger (SSH) model \cite{33,29,43,104,105,106,107} have also been performed, where disorder has been introduced in the hopping to find a non-Hermitian topological Anderson insulating phase \cite{20}. \\

In this paper, we undertake a comprehensive study of the interplay of disorder and symmetry classes in the non-Hermitian SSH model. The point-gap topology in the non-reciprocal SSH model is protected by the sub-lattice symmetry and belongs to Class AIII. We consider a more general model with on-site terms that breaks the sub-lattice symmetry. The resulting system has no symmetry and belongs to Class A, which also has a corresponding point gap topology. In contrast to the Hatano-Nelson model, which also belongs to the same class, we find that in our case, for real on-site terms, there is an extended region in the phase diagram with trivial winding (see Section \ref{sec 3E}).   

We then incorporate disorder, both real and imaginary, to the above mentioned model and numerically study its effects. For convenience of the reader, we summarize our key findings below:

\begin{enumerate}
   \item  We introduce disorder in the on-site potential and the non-reciprocal hopping. For periodic boundary conditions (PBC), we calculate the real-space winding number and the current corresponding to the anomalous chiral mode associated with the spectral topology \cite{48,bessho21,4}. Both the winding number and the chiral current are self-averaging. The value of the chiral current retains its clean system value and is remarkably independent of disorder strength, becoming zero discontinuously when the winding number becomes trivial. This is true for Classes AIII and A.  However, in Class D$^\dagger$ the chiral current is zero on disorder averaging and shows no equivalence to the winding number. 
    
   \item  Next, considering the inverse participation ratio (IPR) and local density of states (LDOS), we show two disorder-induced transitions in the system under PBC. As the disorder strength is tuned, we encounter a mobility-edge phase where a fraction of states are localised, though the complex spectrum has a finite winding. With the disorder-induced jump in the winding number from 1 to 0, the system undergoes a transition from a mobility edge phase to a fully localised phase. The mobility edge in Class A and AIII occurs with respect to real part of the energy, $\Re(E)$, while for Class D$^\dagger$ it occurs as a function of the imaginary part, $\Im(E)$.
    
   \item Under open boundary conditions (OBC), in all three classes A, AIII and D$^\dagger$, upon adding disorder, we again find two disorder-induced transitions characterised by the LDOS. Being a point-gapped topological system, in the clean case, the skin effect is always present. However as the disorder strength is increased, we find a mixed-phase with a fraction of states localised in the bulk. At a critical disorder strength, the skin effect is fully destroyed, with states now localised in the bulk. The real-space winding number computed deep within the bulk also shows a transition to a trivial value at this point. The winding number is a good indicator to track the complete localization of states in the system. 
    
   \item We discover a non-Hermitian Anderson skin effect (NHASE) for all the classes under OBC when we start introducing disorder around the critical line $\delta_c=\pm|1+t_1|$( of the non-disordered system). The winding number shows a direct correspondence to the NHASE, only when all symmetries of the system are broken.
   
\end{enumerate}

The rest of the paper is organized as follows: 
In Section \ref{sec 2}, we describe the non-Hermitian SSH model and discuss its symmetries under different on-site terms. In Section \ref{sec 3}, we briefly summarize some features of the clean system without disorder and present the real space formalism for winding number. Here, we investigate the effects of adding a disorder-free on-site potential and note the changes it causes in the phase diagram. Finally, in Section \ref{sec 4}, we present the numerical results on introducing disorder, using real space calculations of the winding number,chiral current, edge density, IPR and LDOS. Section \ref{sec 5} concludes by summarizing our results and providing an outlook.

\section{The system \label{sec 2}}

\subsection{Our model \label{sec 2A}}

\begin{figure}
    \centering
    \includegraphics[width=0.47\textwidth]{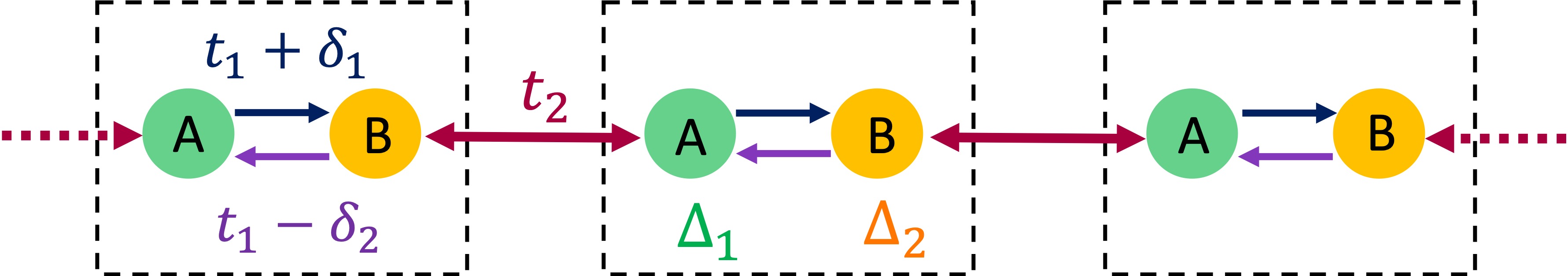}
    \caption{\label{Fig: Schematic} \textbf{Illustrating the model.} Schematic of the non-Hermitian SSH lattice model with hopping and on-site disorder. $A$ and $B$ correspond to the two sublattices of the model while $j$ is the lattice site index. $t_1$ and $t_2$ are the intra and inter cell hopping strengths, respectively. $\delta_{1,2}$ are the non-Hermiticity parameters making the intracell hopping non-reciprocal. $\Delta_1$ and $\Delta_2$ are the on-site potentials added to $A$ and $B$ sublattices, respectively.}
\end{figure}

We consider the generalized non-Hermitian SSH model, which has been widely investigated for its point gap topology \cite{43,44,45,46}. Our generalized Hamiltonian has the form

\begin{equation}
\begin{aligned}
   H & = \sum_j [(t_1 + \delta_1) c_{j,A}^{\dagger} c_{j,B} +(t_1 - \delta_2) c_{j,B}^{\dagger} c_{j,A} \\ 
   & + t_2 (c_{j,B}^{\dagger} c_{j+1,A} + c_{j+1,A}^{\dagger} c_{j,B}) \\ 
  & + \Delta_1 c_{j,A}^{\dagger} c_{j,A} + \Delta_2 c_{j,B}^{\dagger} c_{j,B}]. \label{eq:model}
\end{aligned}
\end{equation}

Here, $c^\dagger$ and $c$ are the creation and annihilation operators, while $t_1$ and $t_2$ are the intra and inter cell hopping strengths, respectively (See Fig.\ref{Fig: Schematic}). The summation over $j$ runs over all sites up to $n$, which is the total number of lattice sites. Special cases of our Hamiltonian are the chiral symmetric ($\Delta_1=\Delta_2=0$) and $PT$ symmetric ($\Delta_1=-\Delta_2=i \mu$) versions of this model, and have been studied in \cite{29,5}. Here we consider, $\delta_{1,2}=\delta+\delta^R_{1/2}$, where $\delta^R_1$ and $\delta^R_2$ are the disorder terms added to the non-Hermiticity parameter $\delta$ -- this can also be thought of as disorder in the forward and backward hopping terms, respectively. Similarly, $\Delta_{1,2}=\Delta_0 + \Delta^R_{1,2}$, where $\Delta_0$ is the disorder-free on-site energy, while $\Delta^R_1$ and $\Delta^R_2$ are the disorder terms added to the on-site potential at $A$ and $B$ sublattices, respectively. For our study, $\delta^R_{1,2}=\delta^R \omega_{\delta_{1,2}}$, where $\delta^R$ is the strength of disorder and $\omega_{\delta_i}$ is chosen randomly from the disorder window $[-1,1]$. Similarly, $\Delta^R_{1,2}=\Delta^R\omega_{\Delta_{1,2}}$, where $\Delta^R$ is the disorder strength and $\omega_{\Delta_i}\in[-1,1]$. We next discuss the symmetries of our model in the absence of disorder.

\subsection{Symmetries\label{sec 2B}}

\renewcommand*\arraystretch{1.5}
\begin{table}
\centering
\caption{\label{Table2} \textbf{Symmetry classes of the SSH model.} Classification of the symmetry classes of the SSH model upon including an on-site potential. Here, SLS refers to sublattice symmetry and $PT$ is parity-time symmetry. The symmetry classes are in accordance with References~\onlinecite{1,18}.}
\begin{tabular}{ccc}
 \hline
 \hline
  On-site potential & Antiunitary symmetry & Symmetry class \\
  \hline
  \hline
 $\Delta_1=\Delta_2=0$ & SLS , $\sigma_z H \sigma_z^{-1}=-H$ & AIII  \\
 $\Delta_1 \neq \Delta_2\neq0$ & None & A  \\
 $\Delta_1=i\mu, \Delta_2=-i\mu$ & $PT$, $\sigma_x H \sigma_x^{-1}=H^*$ & $D^\dagger$ \\
  \hline
\end{tabular}
\end{table}

The non-Hermitian system described by the Hamiltonian in Eq. \ref{eq:model} with $\Delta_{1,2}=0$ possesses a sublattice symmetry (SLS) given by \cite{1} $\sigma_z H \sigma_z^{-1}=-H$ (here $\sigma_{x,y,z}$ are the Pauli spin matrices). Therefore, it belongs to Class AIII.  SLS can be broken by on-site terms. If we consider the non-Hermitian SSH model in momentum ($k$) space, our Hamiltonian with on-site energies $\Delta_{1,2}$ has the matrix form:

\begin{equation}
  H(k)= \begin{pmatrix}
  \Delta_1 & t_2 e^{-ik} +t_1 -\delta_2\\
  t_2 e^{ik} +t_1+\delta_1 & \Delta_2
  \end{pmatrix}.
\end{equation}

For sublattice symmetry we must have $\sigma_z H(k) \sigma_z = -H(k)$. However, for our Hamiltonian,

\begin{equation} 
\begin{split}
\sigma_z H(k) \sigma_z & =  \begin{pmatrix}
  \Delta_1 & -t_2 e^{-ik} -t_1 +\delta_2\\
  -t_2 e^{ik} -t_1-\delta_1 & \Delta_2
  \end{pmatrix}  \\
 & \neq -H(k)
\end{split}
\end{equation}

Hence, any non-zero on-site term breaks the sublattice symmetry of the system. Further, on-site energy can be introduced in several ways. Different choices of the on-site terms and the symmetry classes they correspond to are presented in Table \ref{Table2}. For the present study, we focus on the following types of on-site terms and corresponding disorder:
(a) symmetric on-site disorder: $\Delta^R_1=\Delta^R_2$ in Eq. \ref{eq:model}. Such an onsite term explicitly breaks SLS and places the system in Class A. The symmetric case satisfies the following constraint: $\sigma_x H^T \sigma_x^{-1}=H $, which however is not an antiunitary symmetry.
(b) random on-site disorder: $\Delta^R_1 \neq \Delta^R_2$, which also belongs to Class A.
(c) $PT$-symmetric on-site disorder: $\Delta^R_1=i \mu$ and $\Delta^R_2=-i \mu$. This establishes the symmetry $\sigma_x H \sigma_x^{-1}=H^* $ and belongs to Class D$^\dagger$ \cite{18}.\\

Our calculations show that the sub-lattice symmetric case with only hopping disorder (Class AIII; $\delta^R_{1,2}\neq0,\Delta^R_{1,2}=0$), gives qualitatively similar results as the symmetric on-site disorder case (Class A; $\delta^R_{1,2}=0,\Delta^R_1=\Delta^R_2\neq0$) [for a discussion, see Section \ref{sec 4A}]. Later, we will consider the case of random on-site disorder and $PT$-symmetric disorder to highlight new features of the disordered non-Hermitian SSH system [Section \ref{sec Ddag}, \ref{sec 4D}].\\

\section{Discussions for the Clean System \label{sec 3}}

We first briefly revisit the clean system and discuss some of its important properties. In this case, the disorder parameters in Eq. \ref{eq:model} are set to zero to obtain the usual non-Hermitian SSH model, i.e. $\delta^R_{1,2}=\Delta^R_{1,2}=0$.

\subsection{Energy Spectrum of Clean System\label{sec 3C}}

\begin{figure*}
    \centering
    \includegraphics[width=0.95\textwidth]{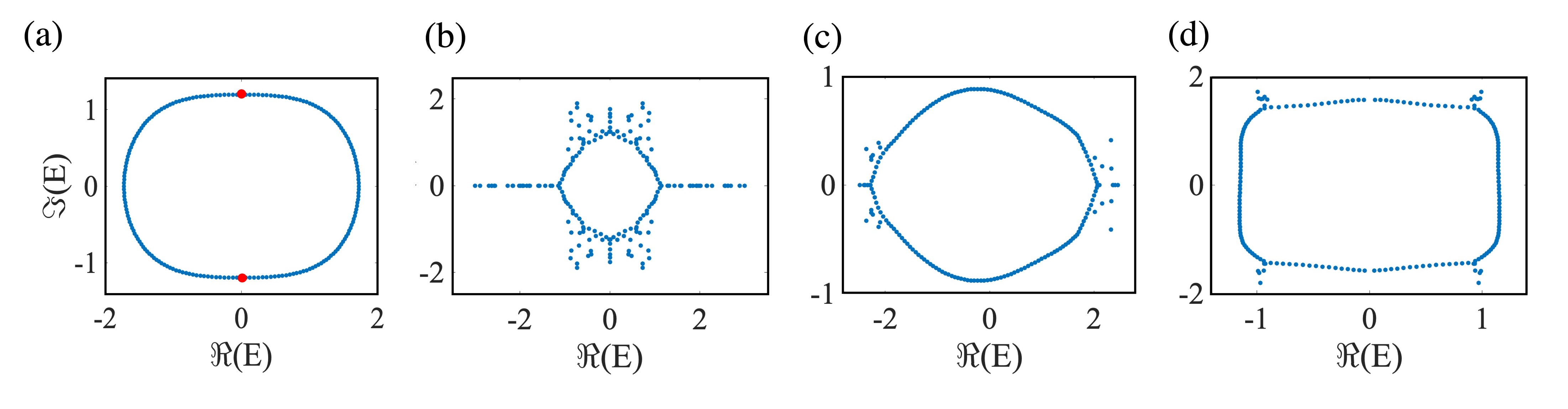}
    \caption{\label{Fig: Energy} \textbf{Non-Hermitian topological phase with disorder belonging to different symmetry classes.} (a) The complex space energy spectrum in the undisordered NH Top phase in the presence of real on-site terms, which has a point gap at $E=0$. In (b), we have added real hopping disorder to the clean system with disorder strength $\delta^R=2.0$. This shows the NH Top spectrum in class AIII. Panel (c) corresponds to class A, where we have added random real on-site disorder to each sublattice with disorder strength $\Delta^R=2.0$. In (d), we have included imaginary on-site disorder of the form $\pm i \Delta^R$ to sublattices A and B, leading to class D$^\dagger$. Despite disorder, the NH Top phase under all the symmetry classes shows a point-gap topology up to a considerable disorder strength. Here we choose $n=200$, $t_1=1.1$, $\delta=1.2$. The chosen parameters correspond to the NH Top phase of the disorder-free system.}
\end{figure*}

The usual complex spectra of the non-Hermitian SSH model without disorder can be classified into four phases \cite{18,61} -- non-Hermitian topological phase (NH Top), Hermitian topological phase (H Top), non-Hermitian trivial phase (NH Triv) and Hermitian trivial phase (H Triv). The different phases can be characterized by the spectral winding number around different base energies. In $k$ space, the spectral winding number is given by \cite{9} 

\begin{equation}
    W= \int_{-\pi}^{\pi} \frac{dk}{2\pi i} \partial_k \: \textup{ln} \: [\textup{det}(H(k)-E_b)], \label{eq:Windingk}
\end{equation}

where $E_b$ is the base energy about which the winding number is calculated and $H(k)$ is the Bloch Hamiltonian of the system in momentum space.
 The winding number is non-trivial in the NH Top phase with $E_b=0$, since at this point the spectrum shows a point-gap (see Fig. \ref{Fig: Energy}a). With non-trivial winding around some complex base energy, there usually occurs a non-zero chiral current in the system which we will investigate in Section \ref{sec 4A}.

\subsection{Winding Number in Real Space\label{sec 3D}}

Let us next summarize the real space formalism for the winding number. First, we define a base energy, $E_b$, around which we wish to calculate the winding. We then construct the matrix $H-E_b I$  and perform a singular value decomposition to get the form $H-E_b I=M S N^{\dagger}$, where $S$ is a diagonal matrix with eigenvalues along the diagonal. Then, we define $Q=M N^{\dagger}$ and $P=N S N^{\dagger}$, such that $H-E_b I=QP$. Now we have obtained the polar decomposed form of the Hamiltonian and we can write \cite{9,10}

\begin{equation}
    W= \frac{1}{L'} \mathrm{Tr'} (Q^{\dagger} [Q,X]). \label{eq:windingreal}
\end{equation}

Here, $X$ is the position operator denoting the position of each lattice site. Before taking the trace, we need to eliminate a sufficient number of lattice sites from both ends of the one-dimensional lattice so that the winding number is determined solely by the bulk properties of the system. So, we define a cut-off length, $l$, from both ends of the one-dimensional lattice. The effective bulk length is then $L'=L-2l$ ($L$ is the total length of the lattice). Here, $\mathrm{Tr'}$, therefore, denotes that the trace has been taken only over this bulk length. This formulation gives a winding number which is quantized and robust, and allows to definitively characterize the different phases and identify phase transitions which may be parameter-driven or disorder-driven. For our calculations we fix $l=0.2 L$.

\subsection{Chiral Modes and Non-Hermitian Skin Effect\label{sec 3A}}

The spectrum of the system under PBC in the non-Hermitian topological phase, shown in Fig. \ref{Fig: Energy}, illustrates that the spectrum accommodates point gaps. 
In such a point-gapped spectrum, the complex energies for which its real part of the spectrum $\Re(E)$ is gap-less  correspondingly have both positive and negative imaginary energy parts. These eigenvalues correspond to dynamically anomalous chiral modes of the system \cite{48,bessho21,4}. Even though gapless modes come in pairs of opposite chiralities as in a Hermitian system, the non-Hermiticity stabilises only one of these modes. The chirality of the modes are determined by the spectral winding and only one of them survive the longest on evolving through time. In order to study the long-time dynamics of the system, it is sufficient to study this chiral mode \cite{34}. For example, in the NH Top phase (Fig. \ref{Fig: Energy}a), the chiral mode is located at $\Re(E)=0$ at the top of the spectrum (shown in red). This chiral mode is responsible for a chiral current, which is unidirectional in nature and is an observable of the system. We develop the formalism to calculate the chiral current in Section \ref{sec 4A}.\\

The non-Hermitian skin effect is another unique property of non-Hermitian systems where, as long as the spectrum accommodates a point-gap topology under PBC, all the eigenstates under OBC migrate to an edge determined by the direction of the persistent current \cite{3,4,49,50,51}. From the spectral topology shown in Fig. \ref{Fig: Energy}, we see that the non-Hermitian phase under different symmetry classes will show a skin effect when we cut open the system under PBC and switch to OBC. Interestingly, the non-Hermitian skin effect can also occur starting from trivial winding and introducing disorder, which we will elucidate later in Section \ref{sec 4D}.

\subsection{Introduction of Real Symmetric On-site Energy\label{sec 3E}}

\begin{figure}
    \centering
    \includegraphics[width=0.45\textwidth]{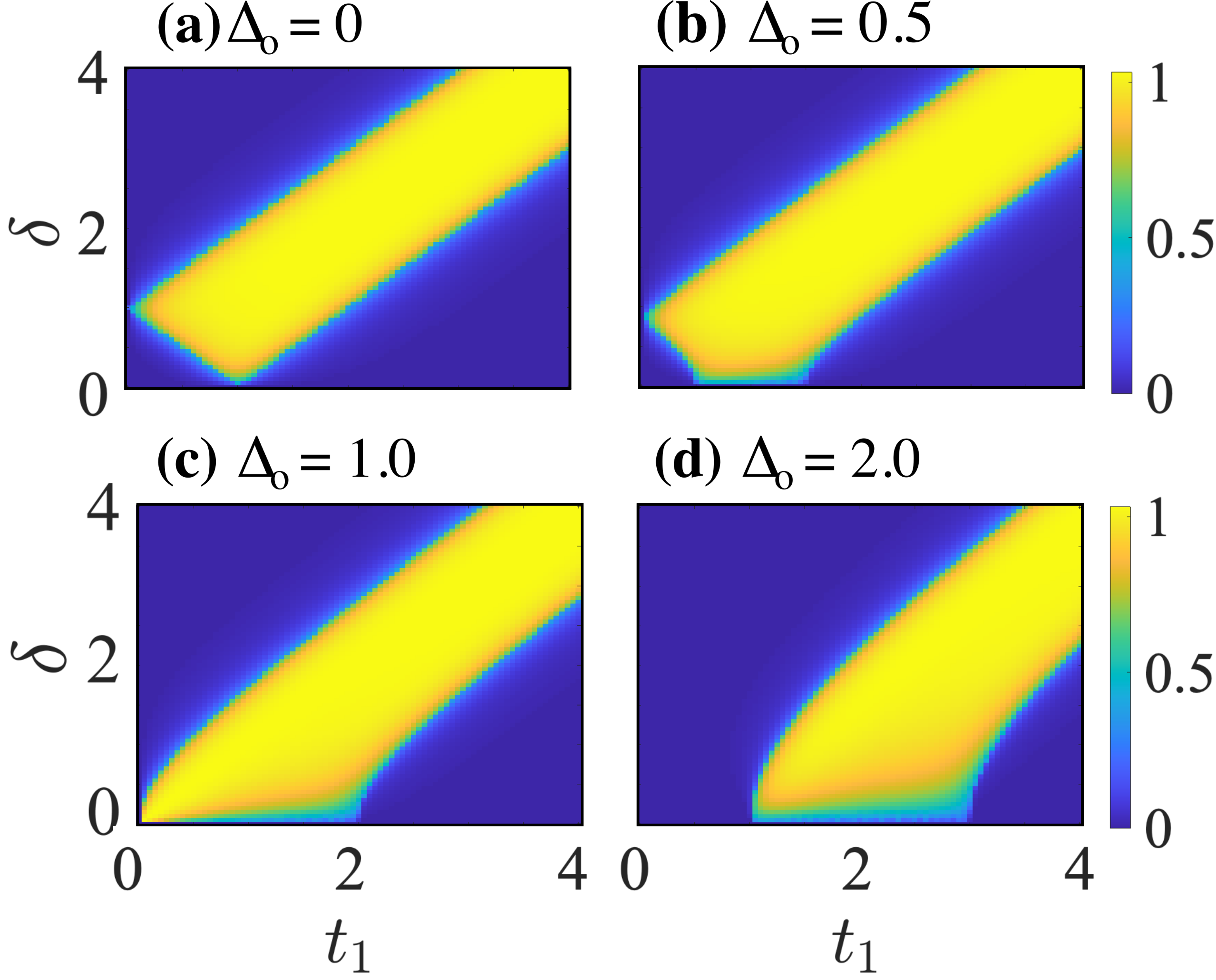}
    \caption{\label{Fig:PurePhase}\textbf{Phase diagram with real symmetric on-site potential.} Winding number as a function of parameters $t_1$ and $\delta$ for different values of on-site energy $\Delta_o$. The critical lines of the regular NH SSH model can be seen in (a) where $\delta_c=\pm|1\pm t_1|$ shows a transition in winding number. As the on-site energy increases (b)-(d) the NH Top region moves towards higher $t_1$ values invading the former H Triv region. Here, $t_2=1$, $n=100$ and base energy $E_b=0$. }
    
\end{figure}

Next, in order to illustrate the effect on of onsite terms on non-Hermitian topological phases we consider the simplest case of  $\Delta_1=\Delta_2=\Delta_o$. Fig. \ref{Fig:PurePhase} shows the evolution of the phase diagram of the system with increasing values of such an on-site energy $\Delta_o$. Fig. \ref{Fig:PurePhase}a shows the well-studied phase diagram \cite{5} of the non-Hermitian SSH model with $\Delta_o=0$, where the non-trivial winding about $E_b=0$ corresponds to the non-Hermitian topological phase. The $t_1=t_2, \delta=0$ is the gapless point in between the two gapped Hermitian SSH topological phases. On addition of even a small non-Hermiticity $\delta$, the spectrum becomes complex and  there is an anisotropy introduced in the right and left movers, effectively resulting in a chiral mode characteristic of the NH Top phase.  On introducing finite $\Delta_0$ at $\delta=0$ (which shifts the `chemical potential' such that it is within a band) one has a window of gapless points around $t_1=t_2$. Now introducing non-Hermiticity $\delta$ over this gapless window leads to the NH Top phase as seen in Fig.\ref{Fig:PurePhase} (b and c). For larger values of $\Delta_o$ the gapless window occurs only at larger values of $t_1$ as seen in \ref{Fig:PurePhase}(d)

\begin{figure}
    \centering
    \includegraphics[width=0.5\textwidth]{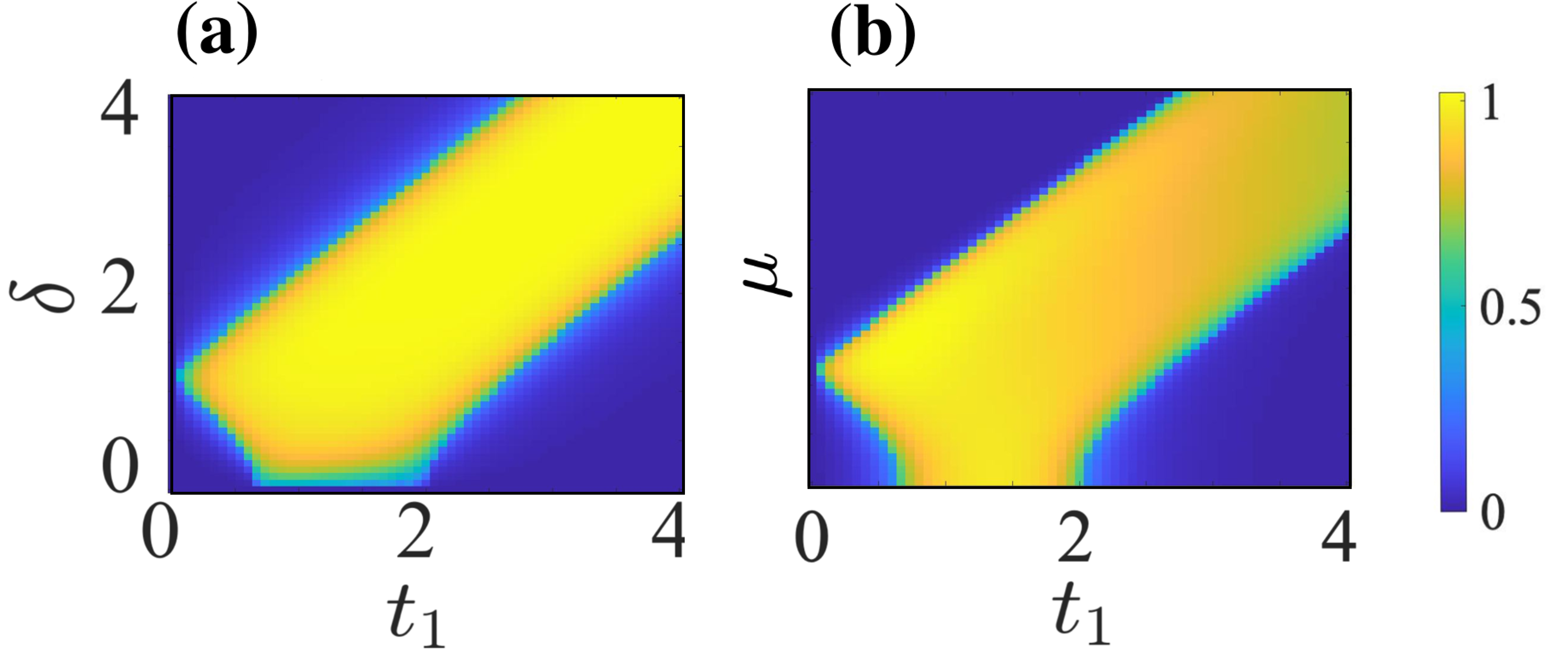}
    \caption{\label{Fig: Wind3} \textbf{Phase diagram with imaginary anti-symmetric on-site potential,  $\Delta_{1,2}=\pm i \mu$}. (a) The winding number as a function of $\delta$ and $t_1$  for $\mu=0.5$. (b) The winding number as a function of $\mu$ and $t_1$ for $\delta=0.5$. The region with non-trivial winding number is the non-Hermitian topological phase which is extended due to the presence of $\mu$. Here $n=100$.}
     
\end{figure}

\subsection{Introduction of Imaginary Antisymmetric On-site Energy\label{sec 3F}}
Here, we look at the changes in the phase diagram on introducing an imaginary antisymmetric on-site energy, $\Delta_1=+i \mu$ and $\Delta_2=-i \mu$, such that the system retains $PT$ symmetry and belongs to the D$^\dagger$ class. The phase diagram is shown in Fig. \ref{Fig: Wind3} and is in agreement with those reported in Reference \onlinecite{18}. Fig. \ref{Fig: Wind3}a shows the winding number as a function of $\delta$ and $t_1$ when $\mu=0.5$ and Fig. \ref{Fig: Wind3}b shows the winding number as a function of $\mu$ and $t_1$ when the non-Hermiticity parameter $\delta=0.5$. The winding number has been calculated with respect to base energy $E_b=0.0$. Hence non-trivial winding number corresponds to the non-Hermitian topological phase. In both cases, we can see an extended region of the same due to the introduction of non-zero $\mu$.

\begin{figure*}
    \centering
    \includegraphics[width=0.95\textwidth]{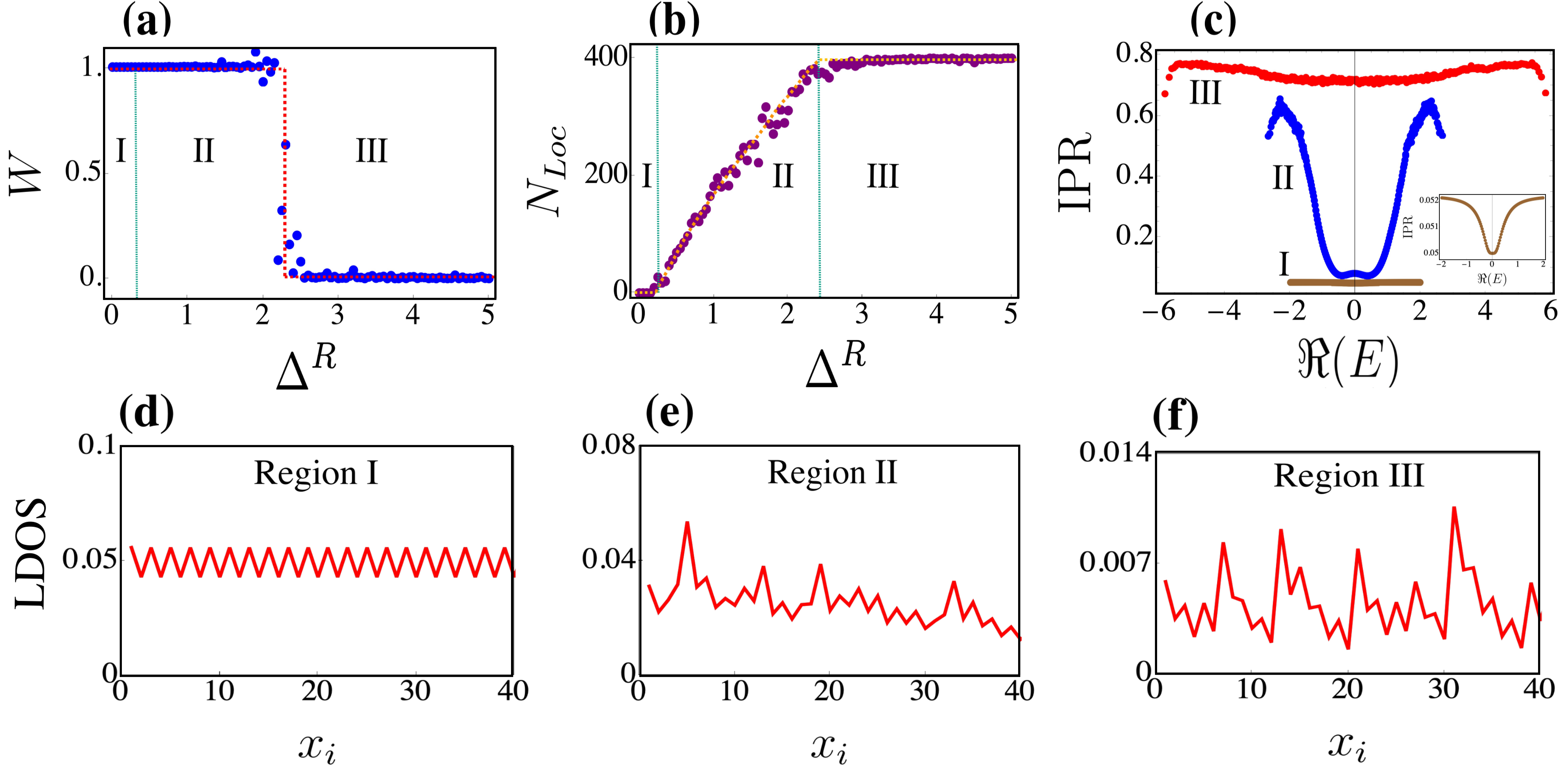}
    \caption{\label{Fig: PBC} \textbf{Characterizing the disorder-induced phases under PBC.} (a) The winding number, $W$, as a function of disorder. (b) The number of localized states (with IPR $\ge$ 0.4) as a function of disorder, demarcating three regions corresponding to the different phases. (c) IPR as a function of $\Re(E)$ for the three regions, where I, II and III correspond to $\Delta^R=$ 0.0, 1.0, 4.5, respectively. The inset shows the zoomed in plot for Region I (in brown). (d) LDOS for Region I ($\Delta^R=0.0$) which is the extended or delocalized phase, (e) Region II ($\Delta^R=1.0$), which is the mobility-edge phase, and (f) Region III ($\Delta^R=4.5$), the bulk-localized phase. All IPR plots have been disorder-averaged. Winding number is self-averaging. LDOS plots (shown only up to the 40th lattice site) are for a representative disorder configuration. Here, number of lattice sites $n=400$, $t_1=1.0$, $t_2=1.0$, $\delta=0.3$, $\Delta_o=0.0$, symmetric on-site-disorder strength $\Delta^R$ varies from 0 to 5 in steps of 0.05.}
\end{figure*}

\section{Introducing Disorder \label{sec 4}}

Next, we introduce disorder in the system through the hopping and the on-site energy terms. In this section, we study how disorder affects various properties such as the chiral current, winding number and density of states. We compare the effect of hopping and on-site disorder and also investigate disorder-driven topological transitions as well as localization transitions.

\subsection{Energy spectra under disorder in the different symmetry classes \label{Disorder_Spectra}}

We find that the point-gap topology of the different phases of the clean system persists even in the presence of disorder for all classes A, AIII and D$^\dagger$, up to considerable disorder strengths. This hints at a persistent non-trivial winding number and a non-zero chiral current even in the presence of disorder which we elaborate in the following sub-sections. In Fig. \ref{Fig: Energy}, we show the non-Hermitian topological phase of the clean system and the changes in the complex space spectra as we introduce disorder transitioning the system between different symmetry classes. Fig. \ref{Fig: Energy}(b) shows class AIII, where disorder has been introduced in the hopping terms, while Fig. \ref{Fig: Energy}(c) shows class A, where a random real on-site energy has been added to each sublattice. Fig. \ref{Fig: Energy}(d) corresponds to class D$^\dagger$, where imaginary disordered on-site energy of the form $\pm i \Delta^R$ has been introduced to sublattices A and B, respectively. In all three disordered cases the spectrum retains its point-gap topology until the critical disorder strength.

\begin{figure*}
    \centering
    \includegraphics[width=0.7\textwidth]{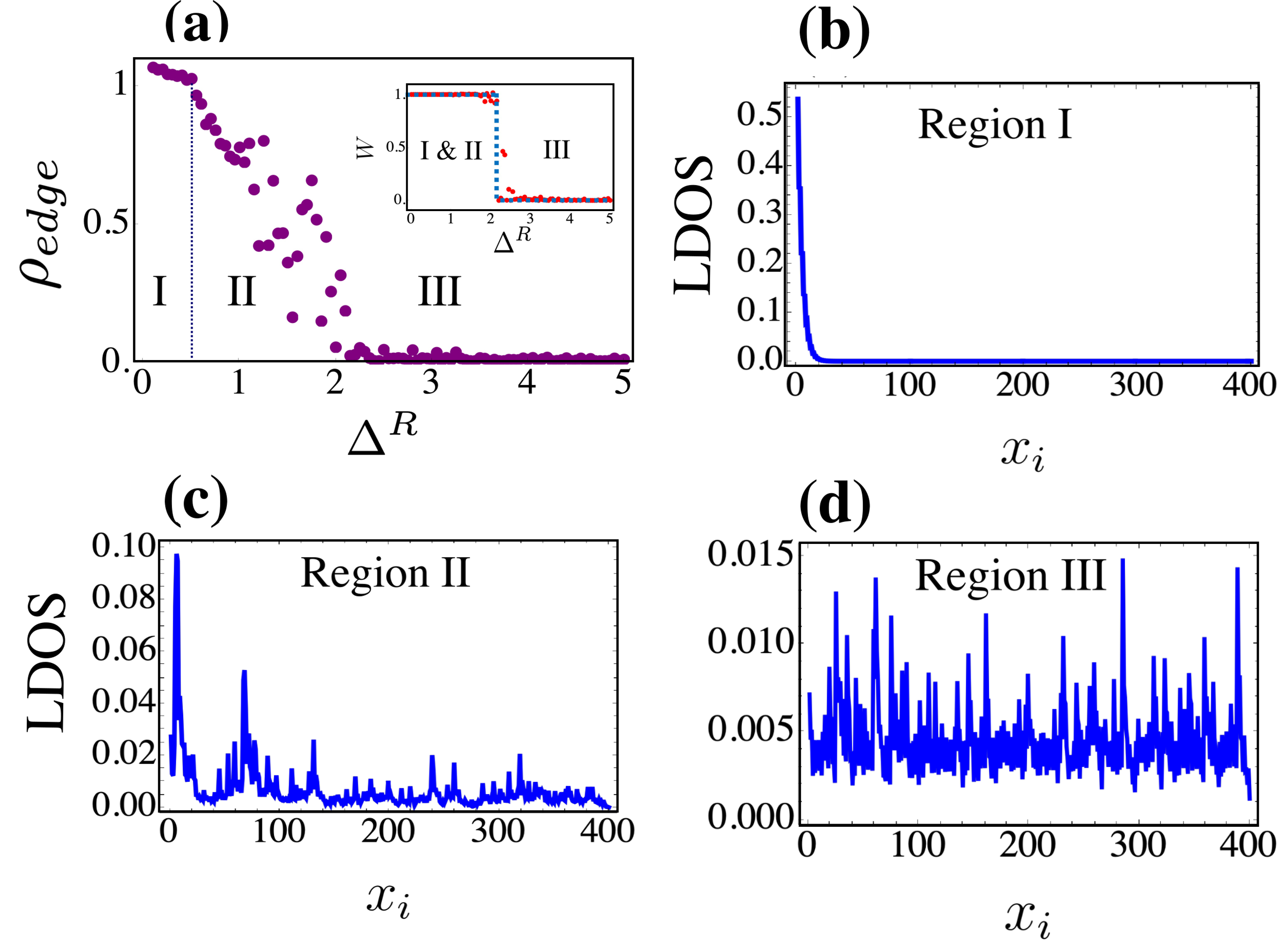}
    \caption{\label{Fig: OBC} \textbf{Disorder induced phases under OBC.} (a) The density of states at the edge ($\rho_{edge}$) which drops down from close to unity as the skin effect disappears. This helps distinguish between Regions I, II and III. The inset shows the change in winding number with disorder which distinguishes between Regions II and III. The LDOS for (b) Region I ($\Delta^R=0.0$), which is the skin phase, (c) Region II ($\Delta^R=1.5$) which is the mixed phase, and (d) Region III ($\Delta^R=4.5$), which is the bulk-localized phase. Here number of lattice sites, $n=400$, $t_1=1.0$, $t_2=1.0$, $\delta=0.3$, $\Delta_o=0.0$ and symmetric on-site-disorder strength $\Delta_1=\Delta_2=\Delta^R$ varies from 0 to 5 in steps of 0.05.}
\end{figure*}

\subsection{Disorder induced localization in Class A and AIII \label{sec 4C}}

The interplay of localization and winding number has been studied widely in the context of disorder and quasiperiodic systems \cite{PhysRevLett.80.5172,PhysRevLett.80.5172,PhysRevB.102.064206,PhysRevB.100.054301,PhysRevResearch.2.033052}.
From our computations, we discover three disorder-induced phases in the system under both PBC and OBC. We characterise these using various measures, such as the winding number, inverse participation ratio (IPR), number of localised states ($N_{Loc}$), the local density of states (LDOS) and the density of states at the edge ($\rho_{edge}$). We briefly describe these quantities in the following: 1) The measure of localization (of a state $\alpha$ with eigenvector $\psi_\alpha$) is calculated using the IPR, which is given by \cite{53} $I_\alpha = \frac{\sum_i |\psi_\alpha(x_i)|^4}{(\sum_i |\psi_\alpha(x_i)|^2)^2}$. $I_\alpha$ close to unity implies that the state is very localized, while a very low IPR is indicative of a delocalized state.  2) To evaluate the LDOS, we calculate $\sum_\alpha |\psi_\alpha(x_i)|^2$ at each lattice site ($x_i$). 3) We characterize a state as localized if its IPR  $\ge 0.4$. The number of such states gives us $N_{Loc}$ which we use to categorize the observed phases. 4) $\rho_{edge}$ gives us the density of states at the edge of the system, where the edge has been quantified as the first two lattice sites.

In Fig. \ref{Fig: PBC}, we present the characterization of the three disorder-induced phases under PBC. We have named them in a manner similar to the disordered phases found in the Hatano-Nelson model \cite{7}. While the winding number is non-trivial, we find the extended (or delocalized) phase (Region I) and the mobility-edge phase (Region II). Region I is characterised by all states being delocalized, hence $N_{Loc}$ is zero (Fig. \ref{Fig: PBC}b). The IPR is very low for all the states (Fig. \ref{Fig: PBC}c) and the LDOS is similar at all lattice sites (Fig. \ref{Fig: PBC}d). Region II is characterized by an intermediate, non-zero $N_{Loc}$ (Fig. \ref{Fig: PBC}b), high IPR for some states beyond a certain $\Re(E)$ and low IPR for other states (Fig. \ref{Fig: PBC}c). The LDOS fluctuates over the lattice sites (Fig. \ref{Fig: PBC}e). In this phase, extended and localized states coexist \cite{68}. Next, as soon as winding number becomes trivial, all the states in the system localize due to high disorder. In this scenario, $N_{Loc} \sim n$ (Fig. \ref{Fig: PBC}a,b). The IPR of all states starts to rise (Fig. \ref{Fig: PBC}c), and the LDOS shows localization of all states in the bulk (Region III) (Fig. \ref{Fig: PBC}f) .\\

Next, we present the characterization of the disorder-induced phases found under OBC in Fig. \ref{Fig: OBC}. Fig. \ref{Fig: OBC}b shows the LDOS for the skin phase, in which all states are localised at the edge of the lattice. Fig.\ref{Fig: OBC}c shows the mixed phase, where a fraction of states show skin effect, while the remaining fraction of states move into the bulk. With a further increase in disorder strength, the system transitions to the bulk-localized phase (Region III), where all the states are localized in the bulk, as can be seen from the LDOS in Fig. \ref{Fig: OBC}d. The IPR is always relatively high as the states are either localized at the edge or localized in the bulk of the system. The three phases can be distinguished with the help of $\rho_{edge}$ and the winding number (Fig. \ref{Fig: OBC}a). $\rho_{edge}$ shows a sharp fall as we enter the mixed phase from the skin phase and subsequently goes to zero in the bulk-localized phase, accompanied by a winding number change from 1 to 0, demarcating Regions II and III.

It is important to note that under both PBC and OBC, a change in winding number coincides with the transition from a partially localized to a fully localized phase. We find that as long as there is a finite non-zero winding, the system does not transition to a fully localized phase. The chosen parameters in the above discussion correspond to class A. We note that class AIII shows the same qualitative features.

\subsection{Chiral Current\label{sec 4A}}

\begin{figure*}
    \centering
    \includegraphics[width=0.7\textwidth]{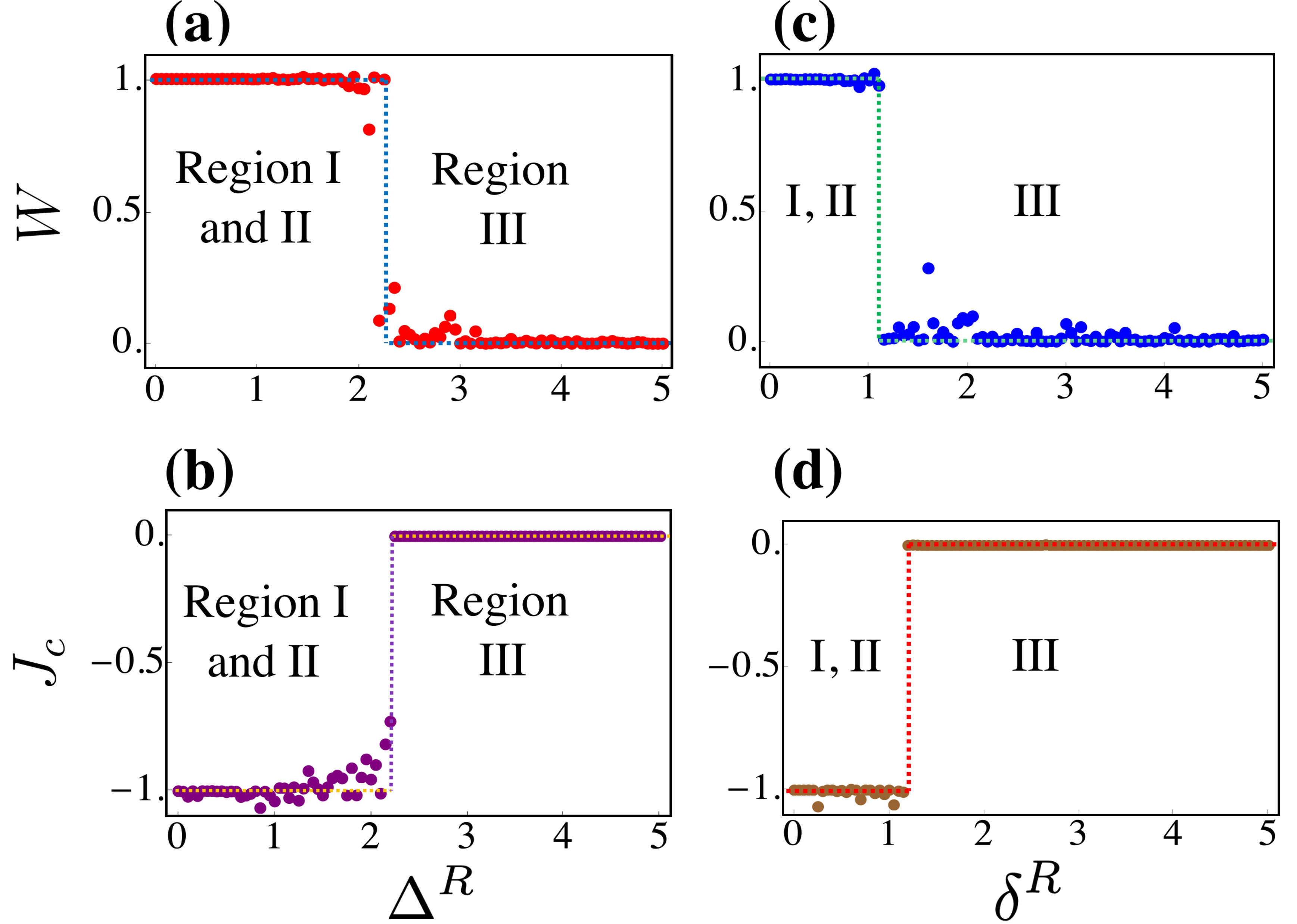}
    \caption{\label{Fig: Current}\textbf{Correspondence between chiral current and winding number.}  (a),(c) The winding number, $W$, as a function of disorder for under PBC. (b),(d) The chiral current as a function of different kinds of disorder in the same system. (a) and (b) correspond to the effect of introducing a symmetric on-site disorder strength (class A) with $\Delta_1=\Delta_2=\Delta^R$. Panels (c) and (d) show the effect of hopping disorder (class AIII). We find that the system is more robust to on-site disorder rather than disorder in hopping and that there is a clear equivalence between winding number and chiral current with disorder. Here $n=400$, $t_1=1.0$, $t_2=1.0$, $\delta=0.3$, and $\Delta_o=0.0$.}
\end{figure*}

As mentioned in Sec. \ref{sec 3A}, the dynamically anomalous chiral mode is on the characteristics of the spectral topology. In our analysis of the disordered system we use the current due to the chiral mode, $J_C$. The current operator is given by \cite{34, 16}

\begin{equation}
    J_{C}= \frac{iL}{2} [\langle c_j^\dagger c_{j+1} \rangle - \langle c_{j+1}^\dagger c_j \rangle], \label{eq:chiralcurrent}
\end{equation}

where the expectation value is taken with respect to the chiral mode, i.e., the eigenvector with the maximum positive imaginary energy at the real-gapless point. While the negative imaginary counter part is also a chiral mode, it decays over long times.
Here, $j\in [1,L-1]$. If $j$ is taken to be odd, $J_C$ corresponds to the intracell current, while an even $j$ yields the intercell current. We note that $J_C$ is independent of the value of $j$ as the same chiral current flows through the entire lattice under PBC.

From Fig. \ref{Fig: Current} we can see that the winding number and the chiral current show a correspondence for both on-site disorder (Fig. \ref{Fig: Current}a,b) and hopping disorder (Fig. \ref{Fig: Current}c,d). These correspond to classes A and AIII, respectively. Regions I, II and III are the same disorder-induced phases found under PBC in Section \ref{sec 4C}.
The chiral current directly corresponds with the winding number and is thus self-averaging i.e without averaging over different disorder configurations, the current approaches a unique mean value for large enough systems. First, let us consider the value of the chiral current in the disorder free case.In momentum space, the Hamiltonian of the disorder-free system can be written as $H(k)=\boldsymbol{d(k)}.\boldsymbol{\sigma}$, where $|\boldsymbol{d(k)}|=\pm \sqrt{t_1^2 +t_2^2 -\delta^2 +2t_1t_2\cos k -2it_2\delta\sin k}=\pm \epsilon(k)$.
Then, the current in the system can be obtained as $j(k)\sim \nabla_k \epsilon(k)$. We obtain

\begin{equation}
    j(k) =\frac{i}{2} (-e^{-i k}/s+e^{i k} s ),
\end{equation}

where, $s=-\sqrt{(t_1+t_2e^{-ik}-\delta)/(t_1+t_2e^{ik}+\delta)}$. We identify the chiral mode at a momentum $k_C$. The chiral current is then $J_C=j(k_C)$. The chiral mode in the non-Hermitian topological phase of the nH-SSH model is  found at $k_C=\pi$ and the value of the current depends on the system parameters we choose.  

Now remarkably, on adding disorder we find that the magnitude of the current is very robust even for substantial disorder strengths up to a critical value where it goes to zero. This is one of the important observations of our work. The symmetry protected anomalous chiral current is self-averaging and is extremely robust to disorder. We also note that our numerical result is consistent with recent field theoretical analysis\cite{23}, which showed that in the continuum limit the  response current at a real-gapless point is exactly given by the winding. 

It is important to note here that we obtain the same qualitative observations irrespective of whether we introduce a disorder in hopping ($\delta^R_{1,2}$; class AIII) or an on-site disorder ($\Delta^R_{1,2}$; class A). However, the non-trivial phase characterized by a non-vanishing chiral current and a finite winding number is more robust to on-site disorder rather than hopping disorder. The hopping disorder has the higher probability of impeding the flow of chiral mode compared to the onsite terms. The chiral current behaves differently in symmetry class D$^\dagger$ as will be discussed in the following subsection.

\subsection{Imaginary On-site Disorder : Class D$^\dagger$\label{sec Ddag}}

\begin{figure*}
    \centering
    \includegraphics[width=0.7\textwidth]{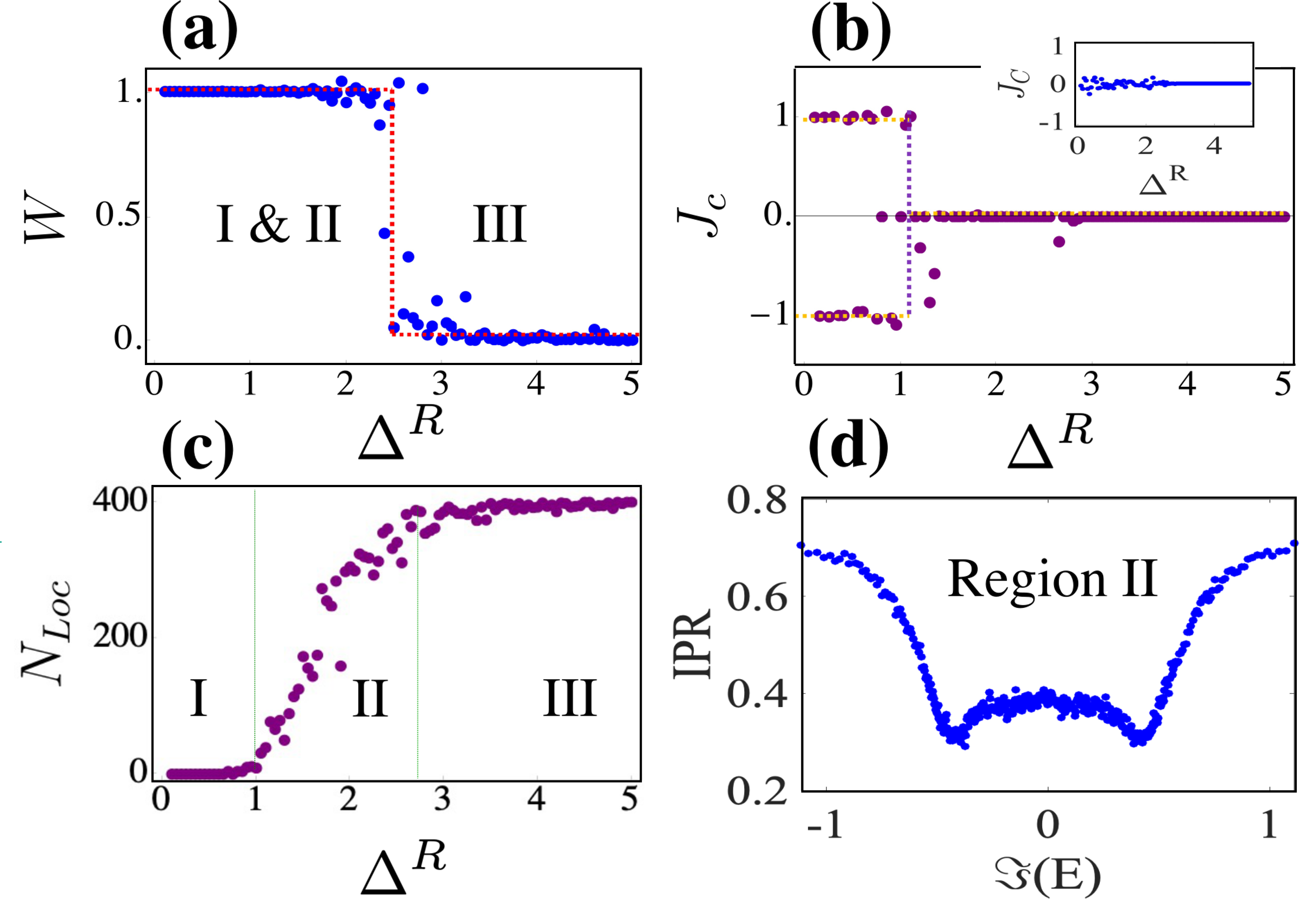}
    \caption{\label{Fig: Ddag_PBC}\textbf{Disorder-induced properties under PBC in Class D$^\dagger$.} (a) The Winding number $W$ as a function of PT-symmetric disorder for the system under PBC. (b) The chiral current as a function of the same disorder which varies from 0 to 5 in steps of 0.05. The chiral current randomly fluctuates between -1 and 1, which when disorder averaged becomes trivially zero (shown in the inset). Panel (c) shows the number of localised states for the same disorder, while panel (d) shows the IPR at $\Delta^R=1.0$ (Region II), which has a mobility edge with respect to the imaginary part of the eigenvalues. The figure shows that for Class D$^\dagger$, winding no. tracks the complete localization in the system demarcating Regions II and III. $N_{Loc}$ helps demarcate between the extended phase (Region I), intermediate mobility edge (Region II) and the bulk localised phase (Region III). The chiral current does not show equivalence to the winding number anymore; rather, it fluctuates between 1 and -1, averaging out to 0 on taking a disorder average. The IPR, contrary to Class A and AIII, shows a mobility edge as a function of $\Im(E)$. Here, no. of lattice sites $n=400, t_1=1.0, t_2=1.0, \delta=0.3, \Delta_o=0.0$.}
\end{figure*}

\begin{figure*}
    \centering
    \includegraphics[width=0.7\textwidth]{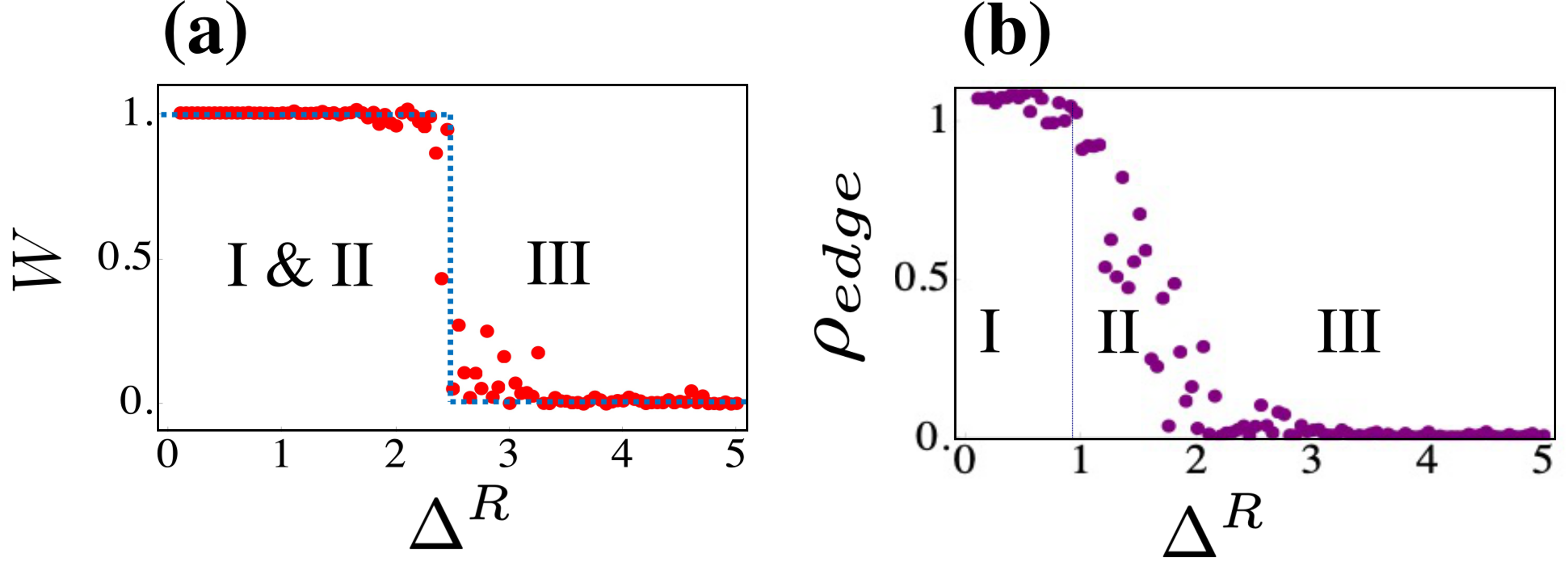}
    \caption{\label{Fig: Ddag_OBC}\textbf{Winding number and edge density under OBC in Class D$^\dagger$.} (a) The winding number $W$ as a function of PT-symmetric disorder for the system under OBC. (b) The density of states at the edge under the same disorder which varies from 0 to 5 in steps of 0.05. The winding number change from 1 to 0 tracks the complete localization of the states into the bulk. The edge density $\rho_{edge}$ shows skin states in Region I, an intermediary mixed phase with partial localization and partial skin states (Region II) and finally Region III where all the states are localised in the bulk. Here, no. of lattice sites $n=400, t_1=1.0, t_2=1.0, \delta=0.3, \Delta_o=0.0$. }
\end{figure*}

In this section, we discuss the effects of introducing an on-site disorder of the form $\Delta_1= i \Delta^R \omega_{\Delta_{1}}$ and $\Delta_2= -i \Delta^R \omega_{\Delta_{2}}$, such that $\omega_{\Delta_{1}}= \omega_{\Delta_{2}}$. This configuration of disorder leads us to symmetry class D$^\dagger$ which, unlike the other classes A and AIII (discussed earlier), includes anti-symmetric imaginary on-site terms chosen randomly from a window of disorder of strength $\Delta^R$. Similar to the previous discussion, here too, we obtain three disorder-induced phases under both PBC and OBC. However, as the disorder is now purely imaginary, different features are found compared to classes A and AIII, which we highlight here.\\

Fig. \ref{Fig: Ddag_PBC} shows the disorder-induced phases under PBC. As before, Region I is the extended phase where the number of localized states, $N_{Loc}=0$ (Fig. \ref{Fig: Ddag_PBC}c), Region II is the mobility-edge phase with an intermediate number of localized states, and Region III is the bulk-localized phase where all the states are localized in the bulk ($N_{Loc}\sim n$). We find that the winding number (see Fig. \ref{Fig: Ddag_PBC}a) tracks the localization physics, showing a sharp transition from 1 to 0 as soon as all the states in the system get localized. Thus, it is effective in demarcating between Regions II and III. Interestingly, the chiral current (shown in Fig. \ref{Fig: Ddag_PBC}b) does not show an equivalence to the winding number anymore, unlike in classes A and AIII. The chiral current fluctuates between 1 and -1 at small values of disorder and at sufficiently large values it stabilises to 0. On disorder-averaging the chiral current over 100 configurations of disorder, we found that $J_C$ is identically zero over the entire disorder region (shown in the inset of Fig. \ref{Fig: Ddag_PBC}b). Thus, in class $D^\dagger$, the chiral current is not robust to disorder nor equivalent to the winding number. Even though the current can be finite for a given disorder configuration, random fluctuations in the equal gain and loss terms destroys the current when disorder averaged.  Turning to localization physics, the mobility edge phase in Region II, is distinct from the mobility edge found for classes A and AIII. Here, the mobility edge occurs with respect to $\Im(E)$ rather than $\Re(E)$, as for the other symmetry classes. This is understandable since the disorder we add here is purely imaginary, contrary to the other symmetry classes. The mobility edge phase is well characterized by the IPR as a function of $\Im(E)$, as shown in Fig. \ref{Fig: Ddag_PBC}d.\\

To characterize the phases under OBC, we use the winding number and the edge density of states ($\rho_{edge}$), as reliable indicators. Fig. \ref{Fig: Ddag_OBC} shows the disorder-induced phases under OBC for class D$^\dagger$. Here, Region I is the skin phase, where $\rho_{edge}\sim1$, i.e., all the states are localized at the skin (Fig. \ref{Fig: Ddag_OBC}b). Region II is the mixed phase with an intermediate edge density, indicating a mixture of skin states and bulk states. Region III is the bulk localized phase, where $\rho_{edge}\sim0$, indicating no skin states. Here too, the winding number tracks the complete localization of the system, clearly demarcating Regions II and III.

\subsection{Disorder around phase boundaries under PBC\label{sec 4B}}

\begin{figure}
    \centering
    \includegraphics[width=0.4\textwidth]{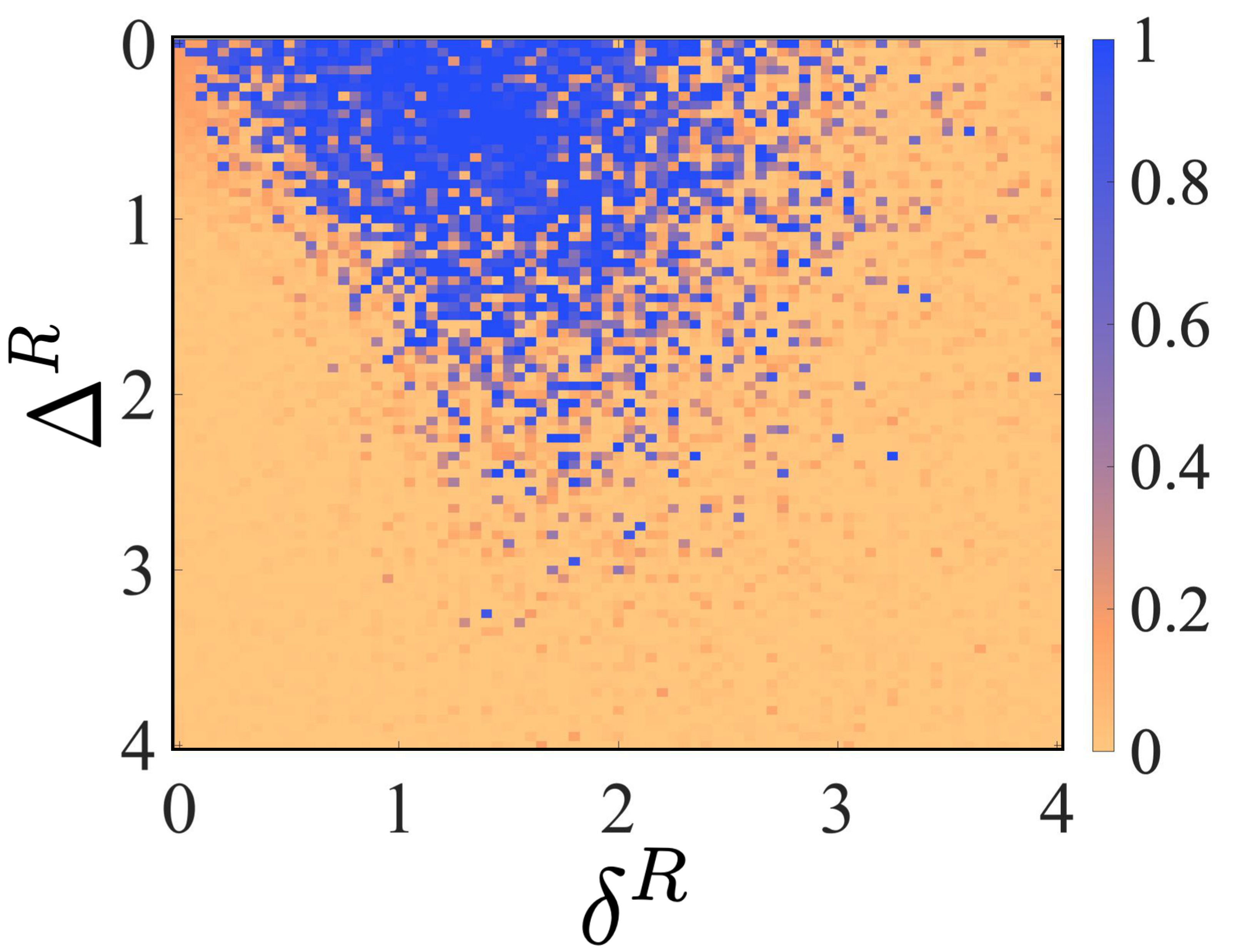}
    \caption{\label{Fig: WPBC} \textbf{Winding number as a function of disorder from a critical line under PBC.} Plot of winding number as function of on-site symmetric disorder and hopping disorder showing a $0 \rightarrow 1 \rightarrow 0$ transition when we sit on the critical line $\delta_c = |1+t_1|$. Here, parameter values $t_1=1.1, \delta_c=2.1, \Delta_o=0.0$ and $n=100$.}
\end{figure}

After investigating the effect of disorder in various classes, next we consider the effect of disorder on a system sitting on the critical lines in the phase diagram of the non-reciprocal SSH model without onsite terms .
In the phases with a non-zero winding number $W$, on the introduction of either hopping ($\delta^R$) or symmetric on-site disorder ($\Delta^R$) or both, $W$ changes from 1 to 0 at a certain parameter-dependent critical disorder strength. Interestingly, starting from the set of critical lines: $\delta_c = \pm |1+t_1|$, demarcating the NH Triv and NH Top phases, $W$ shows a $0 \rightarrow 1 \rightarrow 0$ transition. This behavior persists irrespective of the different disorder configurations and even for choice of parameters slightly displaced from the critical line. Fig. \ref{Fig: WPBC} shows the winding number as a function of disorder strengths $\delta^R$ and $\Delta^R$, when one is positioned on the critical line $\delta_c = |1+t_1|$ under PBC and introduces disorder. As we have established that there exists a correspondence between the winding number and the chiral current for the concerned symmetry classes A and AIII, such a phase diagram indicates that we can start from a system with no chiral current and tune it using either hopping or on-site disorder (or both) to obtain a non-zero chiral current. At the critical lines, $\delta_c = \pm |1-t_1|$, which lie between the NH Top and H Triv phases, $W$ shows a $1 \rightarrow 0$ transition.

\subsection{Non-Hermitian Anderson skin effect\label{sec 4D}}

\begin{figure}
    \centering
    \includegraphics[width=0.4\textwidth]{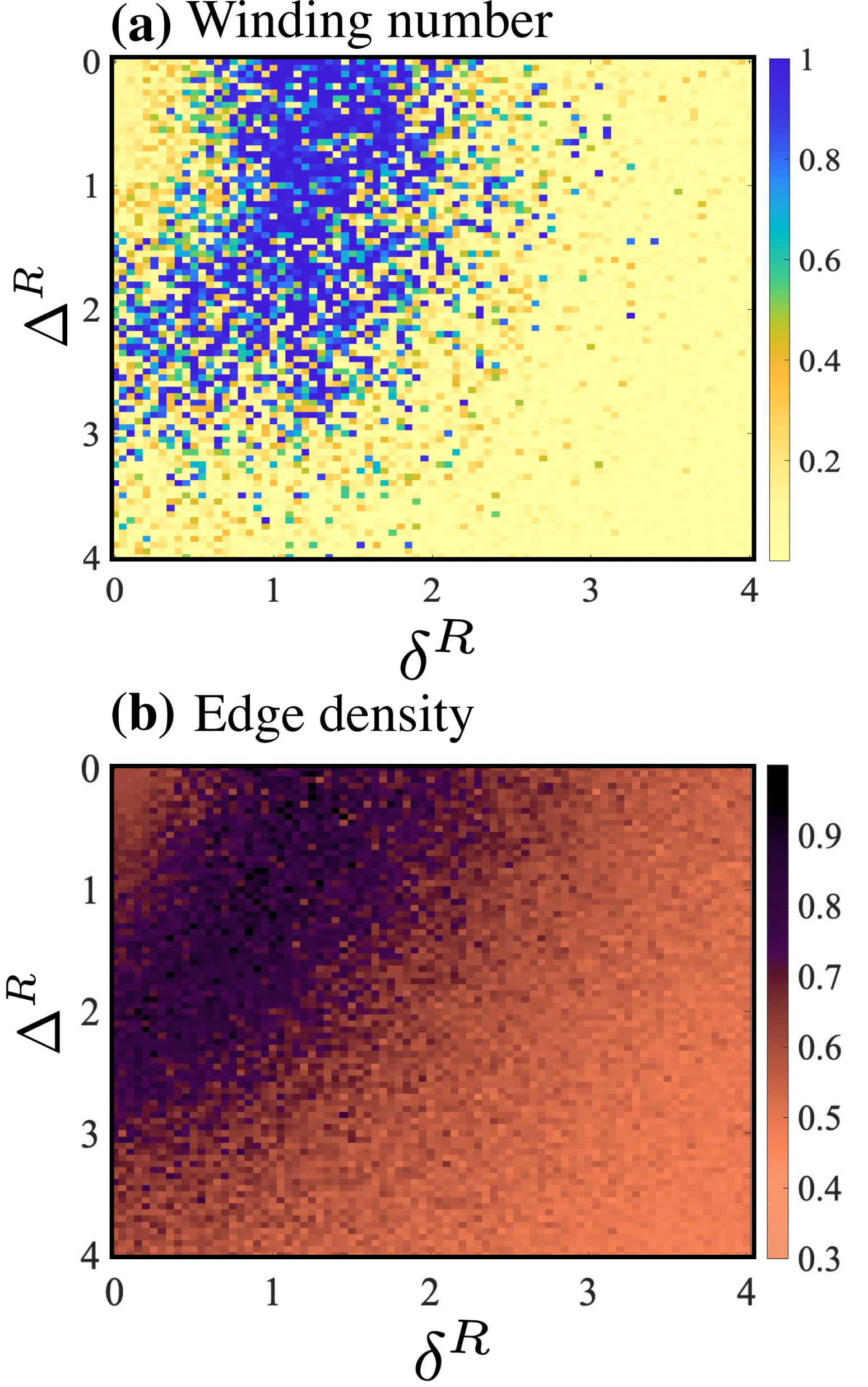}
    \caption{\label{Fig: NHASE} \textbf{The NHASE under OBC.} (a) The winding number as a function of hopping disorder strength $\delta^R$ and random on-site disorder strength $\Delta^R$ which vary from 0 to 4 in steps of 0.05. (b) The density of states at the edge of the lattice as a function of both $\delta^R$ and $\Delta^R$. A system with no skin effect initially develops a skin effect under the effect of disorder which subsequently disappears. Here, $\Delta_{1,2}=\Delta^R\omega_{\Delta_{1,2}}$ such that $\omega_{\Delta_1} \neq \omega_{\Delta_2} \implies \Delta_1 \neq \Delta_2$. When there is no symmetry in the system, the figure shows that the winding number and density of edge states show the same variation with disorder.  Here, $n=100, t_2=1.0, t_1=1.1, \delta=2.1, \Delta_o=0.0$.}
\end{figure}

\begin{figure}
    \centering
    \includegraphics[width=0.4\textwidth]{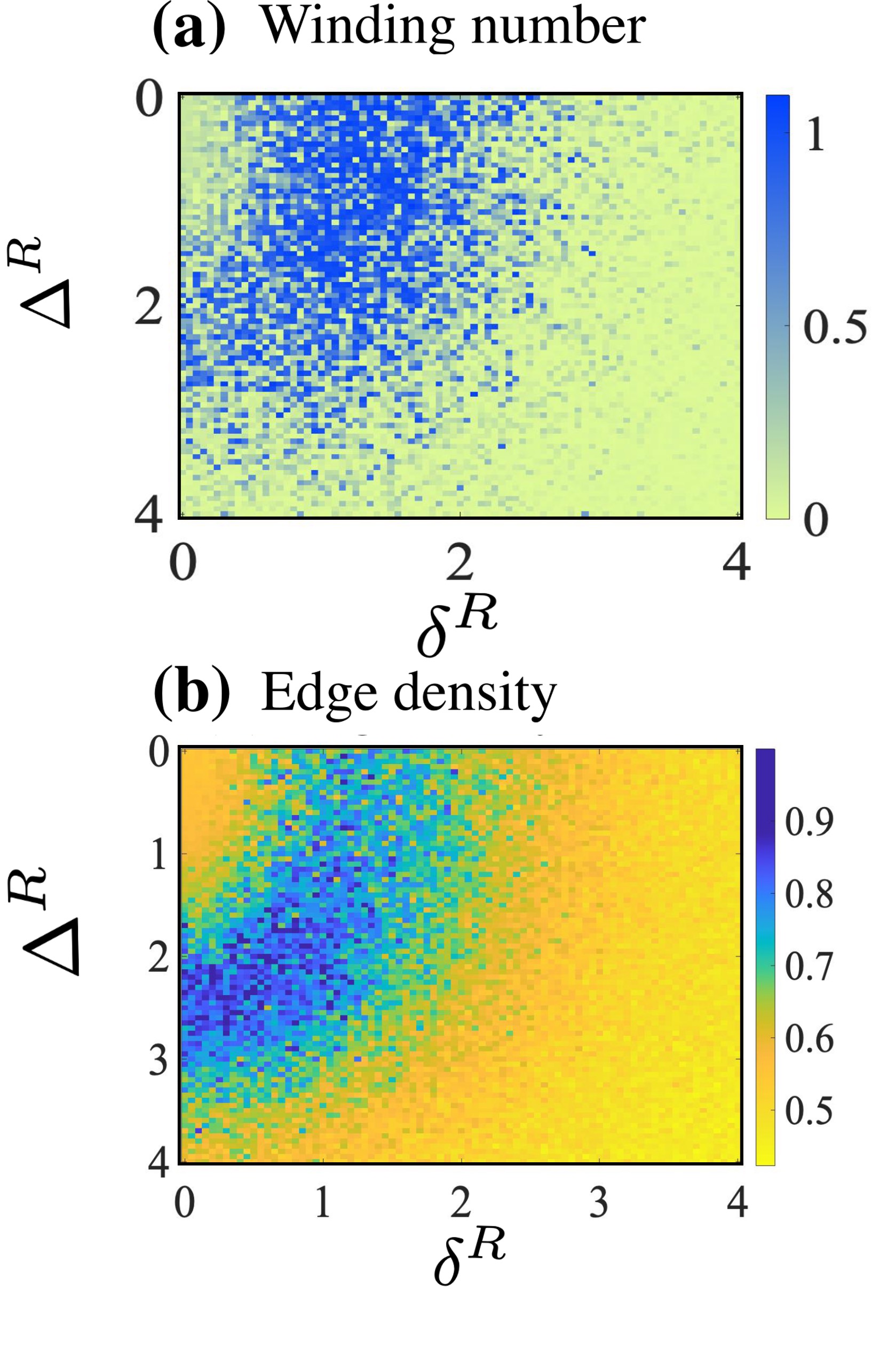}
    \caption{\label{Fig: Ddag_NHASE} \textbf{The winding number and NHASE under OBC in for imaginary disorder.} (a) The Winding number as a function of disorder strength $\delta^R$ for hopping and disorder strength $\Delta^R$ for imaginary on-site terms. Both vary from 0 to 4 in steps of 0.05. (b) The density of states at the edge of the lattice as a function of both $\delta^R$ and $\Delta^R$. Here, $\Delta_{1,2}=\pm i  \Delta^R\omega_{\Delta_{1,2}}$ such that $\omega_{\Delta_1} \neq \omega_{\Delta_2}$. Here too, the winding no. corresponds to the occurrence of anomalous edge states. The parameter values here are: $n=100, t_2=1.0, t_1=1.1, \delta=2.1, \Delta_o=0.0$.}
\end{figure}

Starting at or around any critical point on the line $\delta_c=1+t_1$, under OBC, where there is no skin effect, we find that by introducing disorder in hopping or on-site energy (or both), a skin effect develops and subsequently disappears at high disorder strengths. This disorder-induced appearance of the skin effect has been dubbed the non-Hermitian Anderson skin effect (NHASE) \cite{20}, and was also found recently in the Hatano-Nelson model by Claes and Hughes \cite{9}, upon introducing disorder in the hopping potential

Remarkably, the NHASE in our system occurs irrespective of the presence or absence of symmetries such as SLS, $PT$ or particle-hole symmetry, i.e., the NHASE can be seen for classes A, AIII as well as D$^\dagger$. This anomalous behaviour as a function of disorder is presented in Fig. \ref{Fig: NHASE}(b), which shows that a skin effect occurs at intermediate disorder values and vanishes at large disorder. Here, $\delta^R$ and $\Delta^R$ are the disorder strengths of hopping and on-site real disorder respectively. The same observations hold if we chose $\Delta_1=\Delta_2$ (class A with a constraint) or $\Delta_{1,2}=\pm i \Delta^R\omega$ (class D$^\dagger$) and $\Delta_{1,2}=0$ which corresponds to class AIII. The behaviour of the edge densities in Fig. \ref{Fig: NHASE}(b) and Fig.\ref{Fig: Ddag_NHASE}b are illustrative of this, though their relation to the winding number warrants special attention.\\

In Fig. \ref{Fig: NHASE}(a) and \ref{Fig: Ddag_NHASE}(a), we can see that the winding number shows a similar behavior as the corresponding edge density. This direct correspondence between the winding number and the NHASE occurs only if we break all the symmetries in the system by taking a random on-site disorder, such that $\Delta^R_1 \neq \Delta^R_2$, where $\Delta^R$ could in general be real or imaginary. We show that starting from a system with zero winding number, as we introduce a random disorder in either hopping or on-site energy or both, the system acquires a non-trivial winding at a critical disorder strength, and as soon as the winding number changes to 1, the NHASE develops in the system. Fig. \ref{Fig: NHASE} shows this correspondence between the winding number and the density of states at the edge of the lattice, as we introduce hopping or on-site disorder. The density of states at the edge is the number of states localized at the edge as a fraction of the total number of eigenstates of the system. \\ 
It is important to note that our findings demonstrate that the origin of skin effect in non-Hermitian systems is the point-gap spectral topology under PBC \cite{1}, irrespective of having trivial or non-trivial winding. The occurrence of a non-zero current in the system is a consequence of non-trivial winding   \cite{23}. On the other hand, only when we break SLS ($\sigma_z H \sigma_z^{-1}=-H$), $PT$ symmetry ($\sigma_x H \sigma_x^{-1}=H^*$), can we expect a direct relationship between NHASE and winding, but the NHASE is nevertheless present irrespective of presence or absence of symmetries. The winding number, which is effective in tracking the localization physics in all the symmetry classes A, AIII and D$^\dagger$ is unable to track the Anderson skin effect (NHASE) when there is any underlying symmetry or constraint present in the system.


\section{Summary and Conclusions \label{sec 5}}

In view of studying the interplay between symmetry-protected spectral topology in non-Hermitian systems and disorder, we have considered the addition of both real and imaginary on-site potentials to the non-Hermitian SSH model. We have systematically investigated the effect of different kinds of disorder in different symmetry classes by numerically computing the real-space winding number, localization characteristics and the chiral current associated associated with dynamically anomalous chiral mode.  We found a correspondence between the chiral current and winding number in classes A and AIII. The chiral current is self-averaging under disorder and is remarkably robust, retaining its clean-system-value up to significant disorder strengths and dropping to zero only beyond a critical disorder strength. This robustness of the chiral current may be of value in technological applications. Localization transitions due to disorder have been probed under both PBC and OBC, where the system exhibits three distinct phases. We found that a non-trivial winding persists up to significant disorder strengths, resisting a complete localization of all the states. We have also demonstrated the occurrence of the NHASE in all three symmetry classes (A, AIII and D$^\dagger$), where starting from the disorder-free system, which has no skin effect, the system develops a non-Hermitian skin effect on introducing disorder. One of our striking observations is that the correspondence between the real-space winding number and the NHASE holds only when all symmetries are broken. 

We note that the non-Hermitian effective Hamiltonian is valid for short time-scales before the jump processes occur and the steady state dynamics of non-Hermitian systems requires the consideration of jump operators and a Lindbladian approach. One can incorporate non-Hermitian gain and loss as well as non-reciprocal hopping into the Su-Schrieffer Heeger model using these jump operators as has been discovered in Ref. \onlinecite{PhysRevLett.123.170401}. For systems connected to a suitably generic bath, where the Hamiltonian is quadratic in Fermi operators, the system dynamics can be expressed by a ``quadratic Lindbladian" whose diagonalization gives the complex eigenvalues of the system \cite{PhysRevLett.124.040401}. In such a case, following the Lindbladian formalism one can infer the eigenspectrum for a set of parameter values and hence possibly calculate the winding number which will determine the topological phase. Disorder can be incorporated into the model as a fluctuation in these parameters. A deeper understanding of the effect of disorder, its resilience and the time-dynamics is an interesting direction for future work.

Our formalism is general in terms of the Hamiltonian, winding number, chiral current and the other indicators we use. Hence, we expect our analysis of disorder, non-Hermitian skin effect and the single mode (chiral) current to hold for bosonic cases as well. When considering bosonic or fermionic statistics particularly, one would have to be careful in the way one constructs the many body state. The filling factor and interaction terms would be crucial quantities in the differing physical manifestation of bosons and fermions. Consequences of a bosonic or fermionic system with disorder is a very interesting topic for future studies.

There have already been breakthroughs in experimental realization and probing of non-Hermitian physics in optical photonic and mechanical systems \cite{https://doi.org/10.48550/arxiv.2108.01097,PhysRevX.6.021007,PhysRevApplied.13.014047,Cerjan_2019,Qin_2021,weidemann2020topological,PhysRevB.102.104109,zhang2021observation}. Our results could be corroborated in these settings. On the other hand, further theoretical analyses is required towards understanding the disorder-induced topological transitions, criticality and localization in non-Hermitian systems. This could possibly done under the field theoretical framework of recent works \cite{23,21}. The self-averaging nature of quantities such as the localisation length are understood well in terms of transfer matrices in disordered Hermitian systems. The self-averaging nature of the winding number and the chiral current needs to be addressed in non-Hermitian systems.  Many-body effects in non-Hermitian systems have gained significant attention very recently \cite{34,PhysRevLett.123.090603,PhysRevB.102.064206,PhysRevB.105.165137,Denner_2021}. It would be interesting to study the interplay of symmetry classes, disorder and many-body effects. Another important direction would be studying the different symmetry classes and disorder directly in open quantum systems.

\section*{ACKNOWLEDGEMENTS}
RS would like to thank D. Sen for several useful discussions, and acknowledge S. Chakraborty, A. Banerjee and A. Bandyopadhyay for their help. We thank A. Agarwala and A. Banerjee for a related collaboration. We would especially like to thank Tobias Meng for useful comments on the draft. SSH acknowledges funding from the Deutsche Forschungsgemeinschaft and the National Research Fund Luxembourg via the project TOPREL (ME 4844/3-1 and C20/MS/14764976). SSH would also like to acknowledge funding from the Deutsche Forschungsgemeinschaft via the Emmy Noether Programme ME4844/1-1 (project id 327807255), the Collaborative Research Center SFB 1143 (project id 247310070), and the Cluster of Excellence on Complexity and Topology in Quantum Matter ct.qmat (EXC 2147, project id 390858490). AN acknowledges support from the start-up grant (SG/MHRD-19-0001) of the Indian Institute of Science and from DST-SERB (project number SRG/2020/000153).

\bibliographystyle{apsrev}
\bibliography{Ref}


\end{document}